\documentclass[preprint,jgrga]{agutex2015}
 \usepackage[dvips]{graphicx}
 \newcommand{\rsun}{R_{\odot}}

\authorrunninghead{HATHAWAY AND UPTON}

\titlerunninghead{PREDICTING SOLAR CYCLE 25 WITH SURFACE FLUX TRANSPORT}

\authoraddr{Corresponding author: David H. Hathaway,
Mail Stop 258-5, NASA Ames Research Center, Moffett Field, CA 94035, USA.
(david.hathaway@nasa.gov)}

\begin{document}

%
%

\title{Predicting the Amplitude and Hemispheric Asymmetry of Solar Cycle 25 with Surface Flux Transport}

%
%

\authors{David H. Hathaway,\altaffilmark{1} Lisa A. Upton,\altaffilmark{2}}
  
\altaffiltext{1}{NASA Ames Research Center, Moffett Field, CA 94035, USA.}

\altaffiltext{2}{Space Systems Research Corporation, Alexandria, VA 22314 USA.}

%
%

\keypoints{\item Cycle 25 will be similar in size to Cycle 24
\item Cycle 25 will have a more active southern hemisphere
\item Stochastic variations in the convective flows and active region characteristics limit predictability}

%
%

\begin{abstract}
Evidence strongly indicates that the strength of the
Sun's polar fields near the time of a sunspot cycle minimum determines the
strength of the following solar activity cycle.
We use our Advective Flux Transport (AFT) code, with flows well constrained by observations,
to simulate the evolution of the Sun's polar magnetic fields from early 2016 to the end of
2019 --- near the expected time of Cycle 24/25 minimum.
We run a series of simulations in which the uncertain conditions (convective motion details,
active region tilt, and meridional flow profile) are varied within expected ranges.
We find that the average strength of the polar fields near the end of Cycle 24
will be similar to that measured near the end of Cycle 23,
indicating that Cycle 25 will be similar in strength to the current cycle.
In all cases the polar fields are asymmetric with fields in the south
stronger than those in the north.
This asymmetry would be more pronounced if not for the predicted weakening of
the southern polar fields in late 2016 and through 2017.
After just four years of simulation the variability across our ensemble indicates
an accumulated uncertainty of about 15\%.
This accumulated uncertainty arises from stochastic variations in the convective motion details,
the active region tilt, and changes in the meridional flow profile.
These variations limit the ultimate predictability of the solar cycle.
\end{abstract}

%
%

%

\begin{article}

%
%

\section{Introduction}

The 11-year sunspot cycle represents far more than just a quasi-decadal variation in the number of
sunspots seen on the disk of the Sun.
As the numbers of sunspots wax and wane so do the numbers of x-ray flares, coronal mass ejections,
solar energetic particle events, and geomagnetic storms
\citep[see e.g.][]{Hathaway15}.
Strong space weather events impact our technology, costing us both money and inconvenience.
For example, these sporadic events can disrupt cell phone service and global communication, or even shut down entire power grids.
In addition, the background levels of the Sun's total irradiance, the irradiance
at UV, EUV, and XUV wavelengths, and the flux of galactic cosmic rays also vary with the sunspot number.
The short wavelength irradiance levels alter the Earth's upper atmosphere and ionosphere.
Solar cycle related heating of the thermosphere can change the atmospheric density by an order
of magnitude at spacecraft altitudes (400 km) -- leading to excess satellite drag and the loss
of satellites themselves.
Given these impacts, it is important to provide estimates of future levels of activity
in order to anticipate and/or mitigate the consequences.

Predicting levels of solar activity once a cycle is well underway can now be quite reliable.
Predicting levels of solar activity in a cycle shortly after it starts, or before it starts, is far more difficult.
A wide variety of prediction methods have been used, most with little success.
Comparative studies of prediction techniques \citep{Hathaway_etal99, Petrovay10, Pesnell12}
indicate that precursor methods based on the Sun's polar fields at about the time of cycle minimum have the most merit.
Direct measurements of the polar fields have been used to successfully predict the last four cycles,
Cycle 21 \citep{Schatten_etal78}, Cycle 22 \citep{SchattenSofia87}, Cycle 23 \citep{Schatten_etal96},
and Cycle 24 \citep{Svalgaard_etal05}.
Geomagnetic activity near the time of cycle minimum has been
shown to be a good proxy for the Sun's polar fields \citep{WangSheeley09} and extends
the polar field measurements back 12 sunspot cycles.

\begin{figure}[htbp]
\centerline{\includegraphics[width=20pc]{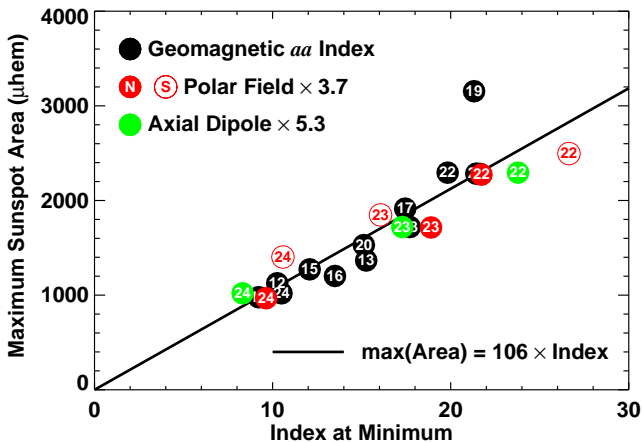}}
\caption{The smoothed maximum of the daily sunspot area in each sunspot cycle is plotted
as a function of three different indicators of the strength of the polar fields at the previous minimum.
The axial dipole moment (scaled by a factor of 5.3) is shown in green.
The polar field above 55\deg\ (scaled by a factor of 3.7) is shown in red.
The minimum in the geomagnetic $aa$ index is shown in black.
All three indicators of the polar fields at cycle minima are very well correlated with the strength of the following cycle maxima.}
\label{fig:PolarFieldPredictors}
\end{figure}

In Figure \ref{fig:PolarFieldPredictors} we plot the strength of the following cycle maximum (as given by
the maximum in the smoothed daily sunspot area) as a function of: 1) the axial magnetic dipole strength at minimum;
2) the average polar field strength poleward of 55\deg\ at minimum; and 3) the minimum in the geomagnetic
{\em aa} index (which occurs at about the time of cycle minimum).
(The smoothing used here is with a Gaussian filter with a FWHM of 24 months convolved with the monthly averages
of the daily values.)
All three indicators of the Sun's polar fields are very well correlated with maximum of the following cycle.
The individual correlation coefficients are 0.99, 0.95, and 0.92 respectively, with a combined correlation
coefficient of 0.90.

This association between the Sun's polar fields and the amplitude of the following cycle is found in most
models for the Sun's magnetic dynamo \citep[see e.g.][]{Charbonneau10}.
In those models the axial dipole field near the time of cycle minimum serves as the background seed field
which is sheared by differential rotation to produce the toroidal field that emerges in active regions.
Dynamo model predictions of Cycle 24 that reset the axial dipole to that observed at the minimum between
Cycle 23 and Cycle 24 were quite successful in predicting the size of Cycle 24 \citep{Choudhuri_etal07, Jiang_etal07}.

Such predictions are, however, sensitive to the timing of the polar field observations.
\citet{MunozJaramillo_etal13B} found that the predictions become less reliable when
polar field observations from well before cycle minima were used.
The success rate falls below 50\% at three years prior to minimum.
(In fact the prediction for Cycle 23 made three years before minimum by \citet{SchattenPesnell93} was significantly off the mark.)

In this paper, we show that reliable forecasts for the amplitude of Solar Cycle 25 can be made even earlier
by  predicting what the polar fields will be during minimum of cycle 24/25, some four years before minimum occurs.

The Sun's polar fields (and its axial dipole) are produced by the emergence of tilted active regions and the transport
of the emerged magnetic flux by the fluid flows in the Sun's surface shear layer \citep{WangSheeley91, Sheeley05, Jiang_etal14}.
Given the emerged active regions and the transport flows (differential rotation, meridional flow, and supergranule convection),
surface flux transport can reproduce the evolution of the polar fields over many cycles \citep{Cameron_etal10}.

Here we use our Advective Flux Transport model (or AFT code) \citep{UptonHathaway14A, UptonHathaway14B},
which combines knowledge of the transport flows along with simulated active region emergence over the next four years,
to predict the polar fields that should be produced on the Sun through the start of the year 2020.
From these predictions, we then estimate the amplitude of Solar Cycle 25.
In addition, we explore the uncertainty in these predictions due to stochastic effects that naturally occur on the Sun
(i.e., convective motions, variability in active region tilt, and variation in the meridional flow).

\section{Surface Flux Transport}

The magnetic flux that emerges in active regions (sunspot groups) is transported across the surface of the Sun
by convective motions (supergranules) and the axisymmetric flows -- differential rotation and meridional flow.
The magnetic elements, with typical field strengths of a kilo-Gauss, are quickly carried to the boundaries of the convection cells
where they form the Sun's magnetic network.
The field in the network is concentrated in downdrafts where it becomes largely vertical.
This vertical (radial) magnetic field is transported by the horizontal flows according to the
flux transport equation:

\begin{equation}
\label{eqn:SFT}
\frac{\partial B}{\partial t}=
   -\frac{1}{\rsun \cos\lambda} \frac{\partial (u B)}{\partial \phi}
   - \frac{1}{\rsun \cos\lambda} \frac{\partial (\cos \lambda\ v B)}{\partial \lambda}+ S
\end{equation}

\noindent where $B = B(\lambda,\phi,t)$ is the radial component of the magnetic field,
$\lambda$ is the latitude, $\phi$ is the longitude, 
$u = u(\lambda,\phi,t)$ is the transport velocity in the longitudinal direction,
$v = v(\lambda,\phi,t)$ is the transport velocity in the latitudinal direction,
$\rsun$ is the solar radius, and $S = S(\lambda,\phi,t)$ is the source term
giving the emergence of flux in active regions.
This equation is derived from the radial component of the magnetic induction equation.

Most surface flux transport modelers approximate the effects of the transport done by the convective motions as a strictly
diffusive process represented by a horizontal diffusion term:

\begin{equation}
\label{eqn:diffusion}
\eta_{H} \left[\frac{1}{\rsun^2 \cos{\lambda}}
                \frac{\partial}{\partial \lambda}\left(\cos\lambda
          \frac{\partial B}{\partial \lambda}\right) +
     \frac{1}{\rsun^2 \cos^2{\lambda}}\frac{\partial^2 B}{\partial
     \phi^2}\right] 
\end{equation}

\noindent where $\eta_{H}$ is a diffusivity.
We model this transport explicitly using an evolving convective flow pattern derived from Doppler measurements of
the spectrum of the convective flows \citep{Hathaway_etal15}.
The convective flow pattern itself is transported by prescribed differential rotation and meridional flow profiles
by solving for the changes in the complex spectral coefficients produced by those axisymmetric flows
\citep[see][]{Hathaway_etal10}.
In addition to this slow evolution of the convection pattern, we give finite lifetimes to the convection cells by
adding small, random rotations to the complex phases of each spectral coefficient at each time step.
The amplitudes of these phase perturbations increase with wavenumber in a manner that reproduces the
decrease in the Pearson correlation coefficient of the Doppler pattern with time.

Our AFT code solves the surface flux transport equation (Equation \ref{eqn:SFT}) using these convective flows along with
the associated differential rotation and meridional flow velocities.
While the solutions for the convective flow velocities are done in spectral space with 4th order Runga-Kutta time differencing,
the solutions for the magnetic field evolution are done in physical space on a grid with 512 equi-spaced latitude points from pole-to-pole
and 1024 equi-spaced longitude points at the equator.
The number of longitude points drops to 512 at $\cos \lambda = 0.5$, to 256 at $\cos \lambda = 0.25$,
and ultimately to 8 just equatorward of each pole,
to keep the actual longitudinal grid spacing (in Mm) close to that used in latitude and in longitude at the equator (4.27 Mm).
Equation \ref{eqn:SFT} is cast in flux conservative finite difference form with second order accurate spatial
differencing and first order Euler time differencing.
We use the convection spectrum of \citet{Hathaway_etal15} out to spherical harmonic degree $\ell = 512$ but with a Hanning taper
on the amplitudes above $\ell = 384$, well beyond the spectral peak at $\ell = 120$ due to supergranules.
The resulting convective motions have maximum velocities on the order of 1500 m s$^{-1}$.
With the first order Euler time differencing, this limits the time steps to about 5 minutes.

Advection equations like Equation \ref{eqn:SFT}  are unstable in the sense that the transported quantity piles up where the flows converge
to produce sharp features, but with Gibbs ringing around the flux concentrations.
We mitigate this problem in the usual manner by adding a diffusion term (Equation \ref{eqn:diffusion}) with a diffusivity just large enough
to minimize the Gibbs ringing.

\section{Flux Transport Flows}

The magnetic flux is transported by the near-surface flows, $u(\lambda,\phi,t)$ and $v(\lambda,\phi,t)$, that include
both the non-axisymmetric convective flows described in the last section and the axisymmetric differential rotation
and meridional flow.
The spectrum of the non-axisymmetric flows is not expected to change much with time and the
differential rotation profile is observed to change only slightly with solar activity.
The meridional flow, however, is observed to vary substantially over the course of a sunspot cycle.
We \citep{HathawayRightmire10, HathawayRightmire11, RightmireUpton_etal12} have measured the
axisymmetric motions of the magnetic elements outside active regions by cross-correlating strips of
data from magnetograms taken at 8-hour intervals from the ESA/NASA {\it Solar and Heliospheric Observatory}
Michelson Doppler Imager (SOHO/MDI) \citep{Scherrer_etal95} and the NASA {\it Solar Dynamics Observatory}
Helioseismic and Magnetic Imager (SDO/HMI) \citep{Scherrer_etal12}.
We have shown \citep{HathawayRightmire11} that this measurement method provides accurate flow
velocities with minimal systematic error.

\begin{figure}[htbp]
\centerline{\includegraphics[width=20pc]{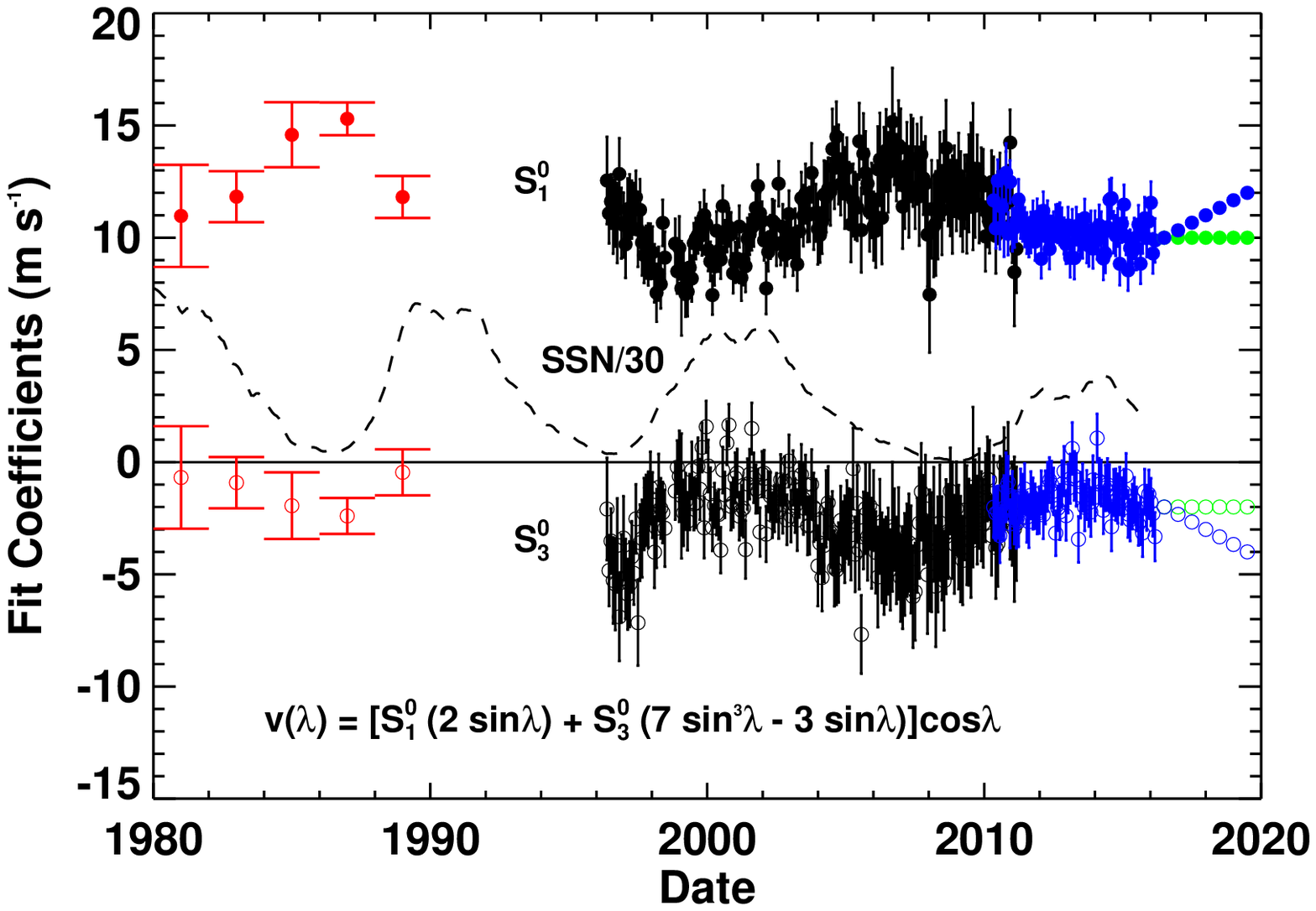}}
\caption{History of coefficients of Legendre polynomials fit to the meridional flow profiles
measured from the motions of the magnetic elements.
The $S_1^0$ coefficients are shown with filled circles.
The $S_3^0$ coefficients are shown with open circles.
Measurements by \citet{Komm_etal93b} in the 1980s are shown in red with 1$\sigma$ error bars.
Measurements by \citet{HathawayRightmire10} from SOHO/MDI are shown in black with 2$\sigma$ error bars.
Updated measurements by \citet{RightmireUpton_etal12} from SDO/HMI are shown in blue with 2$\sigma$ error bars.
Possible future fit coefficients are shown with the lines of dots from 2016 to 2020.
A continuation of the profile measured at the start of 2016 is shown in green.
A projection to a cycle minimum profile in 2020 is shown in blue.
}
\label{fig:MeridionalFlowHistory}
\end{figure}

We fit the differential rotation and meridional flow  profiles with Legendre polynomials and find that the
profiles are well represented with the first 5 (differential rotation) or 6 (meridional flow) polynomials.
In Figure \ref{fig:MeridionalFlowHistory} we show updated values for the coefficients of the two most significant
components of the meridional flow profile.
Earlier measurements by \citet{Komm_etal93b} using a similar measurement technique on ground-based magnetograms
are shown with the red symbols while our measurements from SOHO/MDI are shown in black and measurements
from SDO/HMI are shown in blue.
The meridional flow is fast at sunspot cycle minima (the smoothed sunspot number is shown with the
dashed line in Figure \ref{fig:MeridionalFlowHistory} ) and slow at cycle maxima.
The cycle-related variation in the meridional flow amplitude is about 50\% from cycle maximum to cycle minimum.

\section{The Simulations}

We run a series of simulations in which we evolve the Sun's surface magnetic field starting from an initial magnetic
field map from the end of January 2016 to final magnetic field maps for the
end of December 2019 -- a date expected to be close to the Cycle 24/25 minimum.

The quantities of interest for predicting Cycle 25 are the strengths of polar fields and the axial dipole.
We use historical data from the Wilcox Solar Observatory (WSO) for our comparisons with past cycles so we
adopt, to some extent,  their definitions of these quantities.

The polar fields, $B_N$ and $B_S$, are taken to be the radial field averaged over latitudes poleward of 55\deg, that is

\begin{equation}
\label{eqn:polarFields}
B_N = \int_{55^\circ}^{90^\circ} \int_{0}^{2\pi} B(\lambda,\phi)  d\phi\ \cos\lambda\ d\lambda/ \int_{55^\circ}^{90^\circ} \int_{0}^{2\pi} d\phi\ \cos\lambda\ d\lambda
\end{equation}

\noindent for the north with a similar integral for $B_S$.
Note that the polar fields reported by WSO are for the average line-of-sight field in their northernmost and southernmost pixel.
Those numbers must be divided by a projection factor

\begin{equation}
\label{eqn:projectionFactor}
\int_{55^\circ}^{90^\circ} \int_{-\pi/2}^{\pi/2} \cos\phi\ d\phi\ \cos^2\lambda\ d\lambda/ \int_{55^\circ}^{90^\circ} \int_{-\pi/2}^{\pi/2} d\phi\ \cos\lambda\ d\lambda = 0.2482
\end{equation}

\noindent before they can be compared to the numbers we calculate from our magnetic maps of the radial field using Equation \ref{eqn:polarFields}.

The axial dipole strength, $B_P$, is calculated using the spherical harmonic normalization used at WSO with

\begin{equation}
\label{eqn:dipoleStrength}
B_p = \frac{3}{4\pi} \int_0^{2\pi} \int_{-\pi/2}^{\pi/2} B(\lambda,\phi)  \sin\lambda \cos\lambda\ d\lambda\ d\phi
\end{equation}

\noindent Note that a different normalization ($\sqrt {3/2}$ vs. 3) was used in \citet{UptonHathaway14A}.
The axial dipole strengths shown there must be multiplied by a factor of $\sqrt 6$ to give the same normalization as used here.

The strengths of the polar fields and the axial dipole at the end of the simulations depend upon several factors:
1) the strength and structure of the field in the initial magnetic map,
2) the numbers, strengths, locations, and tilts of new active regions,
3) the strength and structure of the meridional flow, and
4) the diffusive effects of the convective flows.

Our initial magnetic field map is constructed to be a faithful representation of the Sun's surface magnetic field
as measured by MDI/HMI at the end of January 2016.
We have assimilated full-disk magnetograms via Kalman filtering into our AFT code's magnetic maps.
Magnetograms from SOHO/MDI were assimilated every 96 minutes starting in October 1998 (after communications with SOHO were renewed)
and continued with magnetograms from SDO/HMI every 60 minutes starting in May 2010.
Both the polar fields and the axial dipole derived from these maps are closely matched by those reported by the
Wilcox Solar Observatory (WSO) as shown in later figures.
We do find, however, that a calibration coefficient of 1.26 must multiply the WSO polar field values to make them directly
comparable to the MDI/HMI values.

We use the numbers, strengths (area), and locations of active regions from cycle 14, 107 years earlier, as a good
representation of the active regions that will appear in the next four years.
Sunspot Cycle 14 (1901-1913) was similar in amplitude, shape, and hemispheric asymmetry (dominant north early in the cycle
and dominant south late in the cycle) to the current Cycle 24.
In \citet{UptonHathaway14A} we also used active regions from Cycle 14 to successfully predict the reversal
of the axial dipole in Cycle 24 in early 2014 and the strength of the axial dipole in 2016 ($1.4 \pm 0.3$ G predicted vs. 1.2 G observed).

The Royal Greenwich Observatory data for the active regions of Cycle 14 give the longitude, latitude,
and total sunspot area for each active region for each day that the active region was visible.
We use these quantities to give the longitude, latitude, and magnetic flux in  each of the two (balanced)
magnetic polarities comprising the active region.

We convert the total sunspot area, $A$, of the sunspot group into total unsigned magnetic flux of the active region,
$\Phi$, using the relationship given by \citet{Sheeley66} with

\begin{equation}
\label{eqn:area2flux}
\Phi(A) = 7.0 \times 10^{19} A {\rm \ Mx}
\end{equation}

\noindent where the sunspot group area is given in millionths of the area of a solar hemisphere.

We convert the longitude and area of the sunspot group into longitudes for each of the bipolar components using
a longitudinal separation, $\Delta \phi$, between components given by

\begin{equation}
\label{eqn:longitudeExtent}
\Delta \phi(A) = 3^\circ + 8^\circ \tanh (A/500)
\end{equation}

\noindent This is similar to the expression used in \citet{UptonHathaway14A} but is based on measurements
of the centroid positions of the bipolar components of active regions in SOHO/MDI magnetograms.

We convert the latitude and area of the sunspot group into latitudes for each of the bipolar components using
the longitudinal separation given in Equation \ref{eqn:longitudeExtent} and the Joy's Law active region tilt
given by \citet{StenfloKosovichev12} to give a latitudinal separation, $\Delta \lambda$, given by

\begin{equation}
\label{eqn:JoysLaw}
\Delta \lambda(A, \lambda, \Phi) = \Delta \phi(A) \tan(32.\!\!^\circ1 \sin \lambda + \delta \lambda(\Phi))
\end{equation}

\noindent where $\delta \lambda(\Phi)$ is a random variation in active region tilt with a full width
at half-maximum given by \citet{StenfloKosovichev12} as

\begin{equation}
\label{eqn:JoysLawVariation}
\delta \lambda(\Phi) = 25.\!\!^\circ3 + 154.\!\!^\circ7 [1.59/(1.59 + \Phi^{0.84})]
\end{equation}

\noindent with $\Phi$ here given in units of $10^{20}$ Mx.
Note that this variability in active region tilt is a key source of variations in the Sun's polar fields and
axial dipole.
We use Equation \ref{eqn:JoysLawVariation} to produce a series of 8 different realizations for the
tilts of the active regions we assimilate.
This ensemble of realizations allows us to estimate the uncertainty in the predicted quantities.
New active region magnetic flux is added daily provided the area of that active region has increased from
its previous maximum.

The meridional flow profiles are well represented with the Legendre polynomial fit coefficients
shown in Figure \ref{fig:MeridionalFlowHistory}.
We consider two possible future scenarios for the evolution of the meridional flow profile:
1) the meridional flow profile seen at the end of January 2016 continues without changing;
2) the meridional flow profile evolves to more closely match the profile found at the last
cycle minimum.
The fit coefficients for these two scenarios are represented by the symbols in Figure \ref{fig:MeridionalFlowHistory}
that extend from 2016 to 2020 with the green symbols representing the continued profile
and the blue symbols representing the projected (evolving) profile.

\begin{figure}[htbp]
\centerline{\includegraphics[width=20pc]{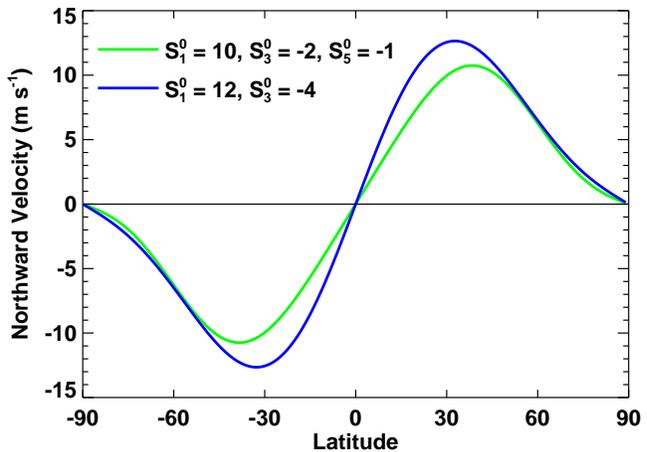}}
\caption{Two possible meridional flow profiles given by the future fit coefficients in Figure 2.
The profile measured at the start of 2016 is shown in green and is referred
to as the continued meridional flow.
The profile projected for 2020 by the coefficient trends is shown in blue and is referred to as the projected meridional flow.
}
\label{fig:MeridionalFlowProfiles}
\end{figure}

The resulting profiles for 2020 are shown in Figure \ref{fig:MeridionalFlowProfiles} using the same
color scheme -- green for continued and blue for projected.
(Note that the primary difference in these profiles is a quenching or weakening of the meridional flow in the active latitudes
in the 2016 continued profile.)
We use each of these meridional flow profile scenarios with the 8 different realizations of the Joy's Law tilt
in the assimilated active regions.

The AFT code allows us to vary the properties of the convective motions (supergranules) that advect the magnetic elements
and produce the magnetic network.
We choose to keep the velocity spectrum and cell lifetimes fixed in all simulations but vary the details of
the convective flows by using 8 different convective patterns produced by shifting the flow pattern by 8
multiples of $45^\circ$ in longitude.
The convective motions are quenched in active regions in a realistic manner by producing a multiplicative mask
for the flows that drops from unity across most of the surface to zero where the magnetic field intensity rises
above a threshold (taken to be 50 G averaged over an area within a radius of 35 Mm for our spatial resolution).
This threshold was determined by examining the masks to see that network and plage fields remained unmasked
while the strong fields in and around sunspots are masked.
This process reproduces the observed decay of active regions as noted by \citet{UgarteUrra_etal15}.
While the convective motions are quenched in active regions, active region magnetic fields are still subject to
differential rotation, meridional flow, and the diffusion imposed to reduce Gibbs ringing.
The 8 different realizations for the convective motions are calculated for both of the meridional flow scenarios. 

\section{Cycle 25 Predictions}

We run the AFT code for 32 different realizations of the Sun's magnetic field evolution over the next four years:
first, the two meridional flow scenarios were run combined with 8 different realizations of Joy's Law tilt in the assimilated active regions,
and then the two meridional flow scenarios were run combined with 8 different realizations of the evolving convective pattern.
In the latter cases Joy's Law was used with no tilt variation.
We save the full surface magnetic field maps at 8-hour intervals and construct magnetic butterfly diagrams
(latitudinal profiles of the magnetic field averaged over longitude and time for each 27-day rotation of the Sun)
from each of the 32 realizations.
We use the magnetic butterfly diagrams to calculate the axial dipole strength and the polar fields averaged
over latitudes above $55^\circ$ for each hemisphere as discussed in the previous section.
This allows us to compare our results for these quantities with those reported from the WSO for earlier cycles.

\begin{figure}[htbp]
\centerline{\includegraphics[width=20pc]{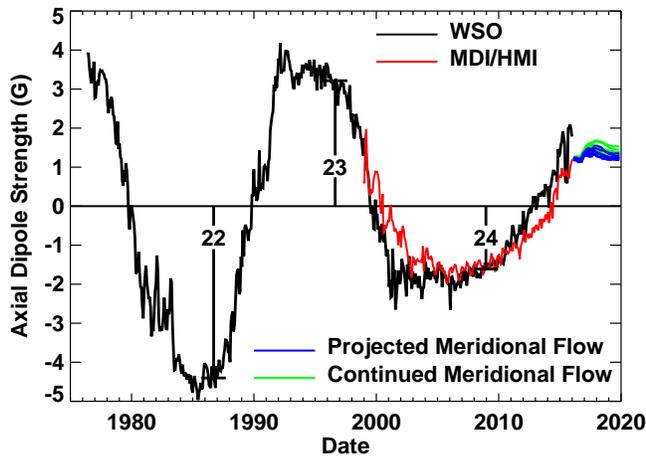}}
\caption{The axial dipole strength predictions with variations in the convective flow pattern and the
meridional flow profile.
The axial dipole strength as reported by the WSO is shown in black.
The axial dipole strength derived from our maps with assimilated MDI/HMI data is shown in red.
The predicted axial dipole strengths obtained with the projected meridional flow profile and eight different
realizations of the convective flow pattern are shown in blue.
The predicted axial dipole strengths obtained with the continued meridional flow profile and eight different
realizations of the convective flow pattern are shown in green.
The axial dipole strengths at the time of sunspot cycle minimum are shown for the last three minima by the
labeled vertical lines.
}
\label{fig:DipoleWithConvectionVariations}
\end{figure}

The axial dipole strength found for the 16 realizations with different meridional flow and convective patterns is
shown in Figure \ref{fig:DipoleWithConvectionVariations}.
This figure includes the WSO measurements (in black) going back to 1976 as well as the MDI/HMI measurements (in red)
starting in late 1998.
Note that at the time of Cycle 24 minimum the axial dipole from MDI/HMI is virtually identical to that from WSO with
no evident need of any multiplicative calibration coefficient.
These results clearly show an axial dipole strength in 2020 similar to that preceding Cycle 24 but
substantially weaker than the strength seen with WSO preceding Cycle 22 and Cycle 23.
While there is some variability in the axial dipole strength due to the variations in the convection pattern,
the systematic variation with the meridional flow profile is more apparent in the offset between
the green and blue lines.
The slower continued meridional flow profile produces a stronger axial dipole.
This behavior of surface flux transport has long been recognized \citep[e.g.][]{Sheeley05}
and is attributed to increased cancellation of opposite leading polarity flux across
the equator when the meridional flow at those latitudes is weaker.  

\begin{figure}[htbp]
\centerline{\includegraphics[width=20pc]{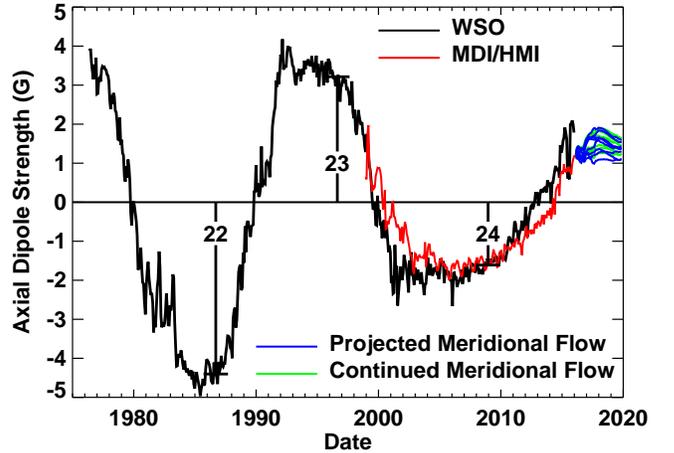}}
\caption{The axial dipole strength predictions with variations in the Joy's Law tilt of active regions and the
meridional flow profile shown in the same manner as in Figure \ref{fig:DipoleWithConvectionVariations}.
}
\label{fig:DipoleWithJoysLawVariations}
\end{figure}

The axial dipole strength found for the 16 realizations with different meridional flow and Joy's Law
tilt is shown in Figure \ref{fig:DipoleWithJoysLawVariations} with the same line colors as in
Figure \ref{fig:DipoleWithConvectionVariations}.
The significant difference with the Joy's Law variations is seen in the larger variation in the axial
dipole strength in 2020 -- large enough to over-power the variations due to
changes in the meridional flow.
The variations in the Joy's Law tilt of active regions clearly has a more substantial
impact on the axial dipole strength.

Here again we see that the axial dipole strength at the start of Cycle 25 (the start of the
year 2020) is similar to to that at the start of Cycle 24 in late 2008 but substantially smaller than
that at the start of Cycle 23 or Cycle 22.
The average of all 32 realizations gives the axial dipole strength at the start of Cycle 25
as $+1.36 \pm 0.20$ G while the WSO gives -1.61 G at the start of Cycle 24,
+3.21 G at the start of Cycle 23, and -4.40 G at the start of Cycle 22.
The 15\% uncertainty in the predicted axial dipole strength is small enough to allow
us to predict that Cycle 25 will be a small cycle like Cycle 24, certainly not as large
as the moderate Cycle 23, and certainly not as small as the cycles in the Maunder Minimum.

We can also address possible hemispheric asymmetry by comparing the average field
intensity above some latitude for each hemisphere.
The latitude chosen is rather arbitrary.
However, to continue our comparisons with the WSO measurements we will use their 55\deg
as the boundary of the polar regions as discussed in Section 4.
Note (as shown in Figure \ref{fig:PolarFieldPredictors}) that these polar fields are good predictors
of the strength (sunspot area) in the associated hemisphere and give the correct sign of the
hemispheric asymmetry in two out of three cycles.

\begin{figure}[htbp]
\centerline{\includegraphics[width=20pc]{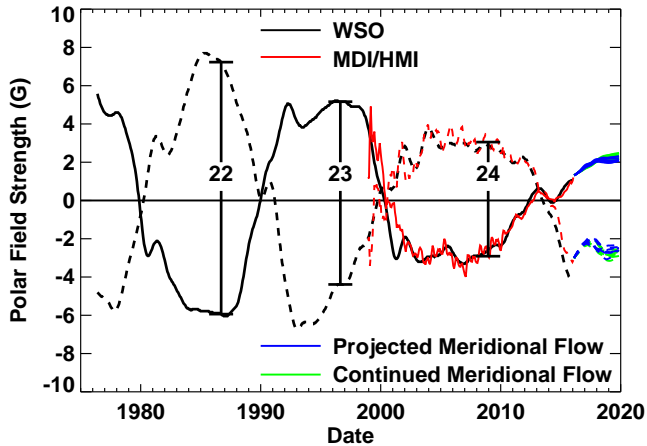}}
\caption{The polar field strength predictions with variations in the convective flow pattern and the
meridional flow profile.
The polar field strength derived from the WSO measurements is shown with a solid black line for the
field in the north above 55\deg latitude and with a dashed line for the field in the south.
The polar field strengths derived from maps with assimilated MDI/HMI data are shown in red.
The predicted polar field strengths obtained with the projected meridional flow profile and eight different
realizations of the convective flow pattern are shown in blue.
The predicted polar field strengths obtained with the continued meridional flow profile and eight different
realizations of the convective flow pattern are shown in green.
The polar field strengths at the time of sunspot cycle minimum are shown for the last three minima by the
labeled vertical lines.
}
\label{fig:PolesWithConvectionVariations}
\end{figure}

\begin{figure}[htbp]
\centerline{\includegraphics[width=20pc]{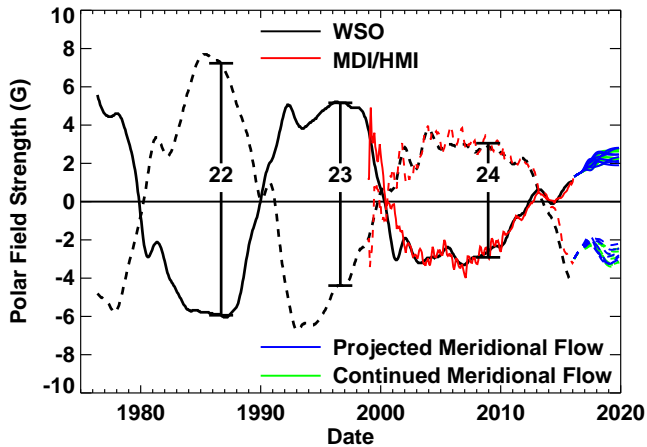}}
\caption{The polar field strength predictions with variations in the Joy's Law tilt of active regions and the
meridional flow profile shown in the same manner as in Figure \ref{fig:PolesWithConvectionVariations}.
}
\label{fig:PolesWithJoysLawVariations}
\end{figure}

The results of the polar field measurements are shown in Figure \ref{fig:PolesWithConvectionVariations}
for the realizations with convective flow variations and in Figure \ref{fig:PolesWithJoysLawVariations}
for the realizations with the Joy's Law tilt variations.
The WSO polar fields are shown in black with a solid line for the northern hemisphere and a dashed
line for the southern hemisphere.
Our polar fields from MDI/HMI are similarly shown in red from October 1998 through January 2016.
The realizations with the continued meridional flow profile are shown in green and the
realizations with the projected meridional flow profile are shown in blue from January 2016 to the start of 2020.
Note that the WSO (multiplied by the calibration coefficient of 1.26) and the MDI/HMI measurements agree very
well where they overlap.

These polar field predictions also indicate that Cycle 25 will be similar in strength to Cycle 24.
The average of the absolute values of the northern and southern polar fields was 2.7 G
at the start of Cycle 24 and is predicted to be $2.5 \pm 0.5$ G at the start of Cycle 25.

If we define the hemispheric asymmetry as

\begin{equation}
\label{eqn:asymmetry}
\frac{|N| - |S|} {0.5(|N| + |S|)}
\end{equation}

\noindent then the predicted asymmetry of Cycle 25 is -0.16 -- the southern hemisphere
should dominate the north.
Note, however, that the predicted asymmetries from the polar fields of Cycles 22, 23, and 24 were
-0.20, +0.16, and -0.11 respectively while the observed asymmetries (in maximum sunspot area) were
-0.09, -0.07, and -0.36.
So this prediction of hemispheric asymmetry should be taken more lightly than the prediction for
the strength of Cycle 25.

Another feature worth noting in Figures \ref{fig:PolesWithConvectionVariations} and \ref{fig:PolesWithJoysLawVariations}
is that we predict that the polar fields in the south will weaken in
late 2016 and into 2017 before recovering.
[Note that these calculations were completed in early 2016.]
This weakening is seen in all of our realizations and is attributed to magnetic field patterns that
are already on the Sun in our initial magnetic map.
The peak in solar activity in early 2014 was followed by a precipitous drop in late 2014.
This drop in activity left low latitude leading polarity flux uncanceled and allowed both polaritiies
to be carried to the poles - higher latitude following polarity first and lower latitude leading polarity later.
This can be seen in all magnetic butterfly diagrams constructed for this time period.
The slow meridional flow and random walk by the convective motions require 1-2 years to
carry that magnetic flux into the polar regions.

\section{Conclusions}

We have used our AFT code to predict the Sun's magnetic field at the start of 2020
as a means of predicting the amplitude and hemispheric asymmetry of Cycle 25 based
on the Sun's polar fields at cycle minimum.
While surface flux transport has been shown to reproduce the magnetic patterns on the
Sun in some detail given knowledge of the active region sources and transport flows,
we do not have detailed knowledge of what those quantities will be in the future.
We do, however, provide knowledgeable estimates and use them, along with their known
variability, to produce a series of 32 realizations for the evolution of the Sun's magnetic field
from the end of January 2016 to the start of January 2020.

We find that the polar fields, as given by the axial dipole strength and the average field
strength above 55\deg, indicate that Cycle 25 will be similar in size to (or slightly smaller than) the current
small cycle, Cycle 24.
We also find (weaker) evidence that the southern hemisphere will be more active than the north.
Small cycles, like Cycle 24, start late and leave behind long cycles with deep extended minima
\citep{Hathaway15}.
We expect a similar deep, extended minimum for the Cycle 24/25 minimum in 2020.

An important result from these simulations of future magnetic fields is the uncertainty
produced by the stochastic variations in both the active region tilt and the convective motions.
While these variations directly limit the accuracy of our polar field predictions to about
15\% after 4 years of simulation, they also suggest that these stochastic variations limit
the predictability of the solar cycle itself.
The accumulated uncertainty over the course of just one sunspot cycle is expected
to make long-range predictions (more than one cycle into the future) very unreliable. 

We note that, while writing this paper, a very similar paper was published by
\citet{Cameron_etal16}.
They also use a surface flux transport code to predict the Sun's axial dipole strength
at the next sunspot cycle minimum and find nearly identical results (although they
refer to the Cycle 24 as a moderate sized cycle when it is clearly smaller
than average).
The primary differences in our methods include: 1) we use detailed convective motions while
they approximate the process with a diffusivity, 2) they use a statistical model for
the emerging active regions while we use a specific example - Cycle 14, and
3) they explore the uncertainty in the initial conditions.
These differences mean that we can determine the uncertainty associated with stochastic
variations in the convection pattern while they can better determine the uncertainties
associated with different locations and sizes of active regions.
We both show the reliability of the method, they on their postdiction of the previous three cycles, we on
our published prediction of the axial dipole over the last four years. 
The fact that we get nearly identical predictions is gratifying.

\begin{acknowledgments}
DHH was supported by a grant from the NASA Heliophysics Grand Challenge Program (NNX14A083G P. Martens, Georgia State University, PI).
LAU was supported by a grant from the NASA Heliophysics Guest Investigator Program (NNH16AC71I I. Ugarte-Urra, George Mason University, PI).
The HMI data used are courtesy of the NASA/SDO and the HMI science team.
The SOHO/MDI project was supported by NASA grant NAG5-10483 to Stanford University.
SOHO is a project of international cooperation between ESA and NASA.
Stanford University operates the WSO with funding currently provided by NASA.
The authors benefited from many useful discussions with Nagi Mansour, Alan Wray, Thomas Hartlep, Robert Cameron, and Jie Jiang,
and from useful comments from an anonymous referee.
The SOHO/MDI magnetograms used in this study are available from: http://soi.stanford.edu/magnetic/index5.html.
The SDO/HMI magnetograms used in this study are available from: http://jsoc.stanford.edu/ajax/lookdata.html.
The WSO data on polar fields and the axial dipole strength are available from: http://wso.stanford.edu.
\end{acknowledgments}

\end{article}

\end{document}